\documentclass[fleqn,10pt,twoside]{article}

\usepackage[english] {babel}

\usepackage {graphicx}
\usepackage {graphics}
\usepackage {floatflt}
\usepackage {latexsym}
\usepackage {caption3} 
\usepackage[labelsep=period]{caption}[2007/11/04 v.3.1e]
\usepackage {amssymb}

\def\noi{\noindent}

\renewcommand{\thesubsubsection}%
        {\arabic{section}.\arabic{subsection}.\arabic{subsubsection}.}

\newcommand{\heads}[2]{\markboth{\protect\small\it #1}{\protect\small\it #2}}

\newcommand{\Arthead}[5]{ \setcounter{page}{#4}\thispagestyle{empty}\noi
    \unitlength=1pt \begin{picture}(500,40)

        \put(0,58){\shortstack[l]{\small\it Gravitation \& Cosmology,
                        \small\rm Vol. #1 (#2), No. #3, pp. #4--#5    \\
        \footnotesize {Proceedings of the 12th Russian Gravitational Conference, Kazan, 20-36 June 2005}    \\
\footnotesize\copyright \ #2 \ Russian Gravitational Society} }

    \end{picture}
	 }     		
\def\prepno#1#2
    {\thispagestyle{empty}
    \noi    \unitlength=1mm
    	\begin{picture}(178,10)
            \put(177,15){\llap{\bf #1}}
            \put(177,10){\llap{\small\rm #2}}
        \end{picture}
          }             

\newcommand{\Title}[1]{\noi {\uppercase{\Large #1}}     }
\newcommand{\Author}[2]{\noi{\large\bf #1}\\[2ex]\noindent{\it #2}   }

\newcommand{\Abstract}[1]{\vskip 2mm \begin{center}
        \parbox{16.4cm}{\small\noi #1} \end{center}\medskip}

\newcommand{\foom}[1]{\protect\footnotemark[#1]}

\newcommand{\email}[2]{\footnotetext[#1]{e-mail: #2}
		\addtocounter{footnote}{1}}

\heads{A.M.Baranov and Z.V.Vlasov}
      {An Exact Interior Extension of a Static Solution for an Electric Charged Ball }

\begin{document}
\twocolumn 
[
\Arthead{11}{2005}{4 ({44})}{315}{316}

\Title{AN EXACT INTERIOR EXTENSION OF A STATIC SOLUTION 

\vspace{.2cm}
FOR AN ELECTRIC CHARGED BALL }

\vspace{.5cm}
   \Author{A.M.Baranov\foom 1 and 
  Z.V.Vlasov\foom 2}   
{\it Dep. of Theoretical Physics,Krasnoyarsk State University,
79 Svobodny Prosp., Krasnoyarsk, 660041, Russia}

\vspace{.3cm}
{\it Received 11 October 2005}
\vspace{.5cm}
\Abstract
{We use the model approach to the description of spherical gravitating static fluid ball with an electric charge in general relativity. The metric is written in Bondi's coordinates. The total energy-momentum tensor (EMT) is chosen as a sum of the EMT of a Pascal perfect fluid and that of the electromagnetic field. An exact solution of the Einstein-Maxwell equations is found, extending a similar solution with parabolic mass density distribution. }
]

\email 1 {alex\_m\_bar@mail.ru}
\email 2 {z\_123@mail.ru}

\section{Introduction}
Models of electric charged stars represent special interest despite the exotic nature of the problem. The problem of finding exact solutions of the Einstein-Maxwell equations does not уне lose its urgency. The electric charge is one of a few quantities which do disappear in stellar collapse. A difficulty of finding exact solutions is connected with the nonlinearity of the gravitational equations. 

The Reissner-Nordstrem solution (see [1]) is an exact exterior solution of the Einstein-Maxwell equations. It describes the exterior field of a static charged star. But the problem of finding internal solutions of Einstein-Maxwell equations is harder. One exact solution of this problem is presented in [2], [3]. It describes a charged perfect Pascal fluid with a parabolic distribution of mass density, but without introducting a specific equation of state. 

In the present article, we present an extention of the model of [2], [3], connected with an extension of the mass and charge density distributions. A new exact static solution for a charged fluid ball is obtained.

\section{The metric and the Einstein-Maxwell equations}

We consider an interior static star model as a ball filled with an electrically charged fluid, using the Einstein-Maxwell equations. The metric is
$$
ds^2 = F(r)dt^2 +2L(r)dtdr-r^2(d{\theta}^2 + sin{\theta}^2d{\varphi}^2)
\eqno{(1)}
$$ 
with radial $(r)$ and angular $(\theta,\,$ $\varphi)$ variables. The vacuum velocity of light is chosen to be equal to unity. 

The interior energy-momentum tensor (EMT) is a sum of the perfect 
fluid and electromagnetic EMTs:
$$
T_{\alpha \beta}=T_{\alpha \beta}^{fluid}+T_{\alpha \beta}^{em},
\eqno{(2)}
$$
where
$$
T_{\alpha \beta}^{fluid}= (\mu +p)\,u_{\alpha}\,u_{\beta} -p\,g_{\alpha\,\beta}; 
$$
$$
T_{\alpha\beta}^{em}=\frac{1}{4\pi}\left(-F_{\alpha\sigma}F_{\beta}^{\sigma}+\frac{1}{4}_{\alpha\beta}F_{\sigma\tau}F^{\sigma\tau}\right)
\eqno{(3)}
$$
and $F_{\alpha\beta}$ is the electromagnetic field tensor.

The Einstein-Maxwell equations are
$$
  (F/xL^2) (ln L)^{\prime} = \chi (p+\mu);
\eqno{(4)}
$$
$$
 (F/xL^2) (ln L)^{\prime} - (1/2L^2)(F^{\prime\prime} +2{F^{\prime}}/x
- {F^{\prime}}(ln L)^{\prime}) = 
$$
$$
= -\chi (p + W_{el});
\eqno{(5)}
$$
$$
(1/x^2)(-1+(F/L^2)+(xF^{\prime}/L^2 -{xF (ln L)^{\prime}}/L^2) = 
$$
$$
= -\chi ((\mu -p)/2 + W_{el});
\eqno{(6)}
$$
$$
  (x^2F/L)^{\prime} = 4\pi R \rho x^2 L/\sqrt{F},
\eqno{(7)}
$$
where $\prime = d/dx,\,$ $x= r/R $, $0 \leq x \leq 1 $,
$R$ is the stellar radius, $ \mu(x) $ is the mass density, $p(x)$ is the pressure.
$W_{el} = {E^2}/8\pi L^2$ is the observable electric field energy density, 
$E$ is the electric field strength, $\;\rho(x)$ is the electric charge density, $\,\chi = \varkappa\,R^2 = 8\pi\,R^2.$

After the replacement $\varepsilon = F/L^2 $ and with the new variable 
$$ 
d \zeta= \displaystyle\frac{x dx}{\sqrt{\varepsilon(x)}} = 
\displaystyle\frac{x dx}{\sqrt{1- \Phi(x)}}
\eqno{(8)}
$$ 
the set of the Einstein-Maxwell equations 
is reduced to the nonlinear spatial oscilartor equation
$$
G^{\prime \prime}_{\zeta \zeta} + {\Omega^2(\zeta(x))} G = 0,
\eqno{(9)}
$$

\noindent
written in terms ofthe new variable $ \zeta$ and $\;G=\sqrt{F} = \sqrt{g_{00}}$.
Here 
$$
\Omega^2 = - \displaystyle\frac{d}{dy}\left(\frac{\Phi}{y}\right)- \displaystyle\frac{2\chi}{y} W_{el},
\eqno{(10)}
$$
where the function $\Phi(x)\,$ plays the role of Newton's gravitational potential inside the star, 
$$
\Phi=1-\varepsilon=\frac{\chi}{x}\int(\mu(x)+W_{el}(x))\,x^2 dx.
\eqno{(11)}
$$
The pressure can be written as
$$
\chi p = \chi W_{el}-\displaystyle{\frac{\Phi}{x^2}+\frac{1}{x}}(1-\Phi)(\ln F)^{\prime}.
\eqno{(12)}
$$

In [2], [3], the mass density distribution is taken as parabolic function
$$
\mu(x) = \mu_0(1-b_0\,x^2),
\eqno{(13)}
$$
where $b_0=(\mu_0-\mu_R)/\mu_0 \leq 1;\; \mu_0=\mu(x=0); \;\mu_R=\mu(x=1).$ 

The energy density of the electric field is taken in [2], [3] by analogy to classical electrostatics:
$$
\varkappa W_{el} = \lambda_0^2\, x^2
\eqno{(14)}
$$
with $\lambda_0=const,$ and electric charge density  was found in form
$$
\rho(x) = \rho_0 \,\sqrt{1-\Phi(x)},
\eqno{(15)}
$$ 
where $\rho_0 = \rho(x=0).$

If we take the mass density as the function 
$$
\tilde{\mu}(x) = \mu_0 (1-b\,x^2)^3
\eqno{(16)}
$$
and the electric field energy density as
$$
\varkappa W_{el} = \lambda(x)^2 x^2
\eqno{(17)}
$$
with 
$$\lambda(x) = (4\pi R \rho_0/3)(1 - 3ax^2/5),
\eqno{(18)}
$$ 
$b =1 - (\tilde{\mu}(x=1)/\mu_0)^{1/3} = const;\,$ $a= 80b/63\;$ (as an extension of the result of [2], [3]), we obtain an exact static solution of the Einstein-Maxwell equations [4]
$$
G(x) = G_0 \cos(\Omega_0 \zeta(x) + \alpha)
\eqno{(19)}
$$
as the solution of the harmonic spatial oscillator equation
$$
G^{\prime \prime}_{\zeta \zeta} +  \Omega_0^2 G = 0
\eqno{(20)}
$$
with the new electric charge density 
$$
\tilde{\rho}(x) = \rho_0 (1-ax^2)\sqrt{1 - \tilde{\Phi}(x)},
\eqno{(21)}
$$ 
where 
$$
a = 1-\displaystyle\frac{\rho(x=1)}{\rho_0\,\sqrt{1-\tilde{\Phi}(x=1)}} = const,
\eqno{(22)}
$$
and here
$$
\Omega_0^2 = \displaystyle\frac{\chi}{60^2}(1701\,\mu_0\,a-1760\,\pi\,\rho_0^2\,R^2).
\eqno{(23)}
$$ 

$$ 
\tilde{\Phi}(x=1) = \eta_0 -\displaystyle\frac{Q^2}{R^2} \equiv \eta^{*}, 
\eqno{(24)}
$$
where $\,Q\,$ is the integral electric charge of the star; $\,\eta_0 = 2m/R$ is the compactness of the stellar model without an integral electric charge, and $m$ is the integral mass of charged fluid ball.

The general expression for the metric function $g_{00} = F = G^2$ is
$$
F(x)=G_{0}^2 \,cos^2(\Omega_{0}\cdot\zeta(x)+\alpha_0)
\eqno{(25)}
$$
and the function $g_{01} = L$ can be found easily from $L = (F/(1-\tilde{\Phi}))^{1/2}.$

Here the Reissner-Nordstrem solution is the exterior solution.
From the boundary conditions we have 
$$
G_0 = \left( 1- \eta^{*}+ \displaystyle\frac{2 \eta^{*}-\eta}{2 \Omega_0}\right)^{1/2}
\eqno{(26)}
$$
and
$$
\tan(\Omega_0\,\zeta(x=1)+\alpha_0)= 
-\displaystyle\frac{2 \eta^{*}-\eta}{2 \Omega_0 \sqrt{1-\eta^{*}}}.
\eqno{(27)}
$$

\section{Summary}

This article demonstrates the Einstein-Maxwell's equations reduction in Bondi's coordinates to the equation of a nonlinear spatial oscillator. The source of gravitational field is a perfect electrically charged Pascal fluid. The assumption on the behavior of the mass-energy density and that of the of electric field was used here. This solution is an extension of the solution found earlier in [2], [3].


\small
\begin{thebibliography}{99}

\bibitem{1}
      A.S.Eddington, "The mathematical theory of relativity", Cambridge, 1924.
\bibitem{2}
      A.M.Baranov, Dep. in VINITI USSR, No.6729-73.
\bibitem{3}
     A.M.Baranov, {\it Vestnik of Krasnoyarsk State University. Phys. Math. Sci.\/}, $\,$ No.1, 5 (2002).
\bibitem{4}
     A.M.Baranov, Z.V.Vlasov {\it Vestnik of Krasnoyarsk State University. Phys.
 Math. Sci.\/},$\,$ No.1, 4 (2005).

\end{thebibliography}
\end{document}